\documentclass[aps,groupedaddress,showpacs,%
amsmath,preprintnumbers,12pt]{revtex4}
\input{amssym}
\usepackage{hyperref}

\def\be {\begin{equation}}
\def\ee  {\end{equation}}
\def\bea {\begin{eqnarray}}
\def\eea {\end{eqnarray}}
\def\nn {\nonumber}
\begin{document}
\preprint{ }
\title{Quantum Hamiltonian for gravitational collapse}
\author{Viqar Husain  and Oliver Winkler}
\email{vhusain@unb.ca, owinkler@perimeterinstitute.ca}
\affiliation{ Perimeter Institute for Theoretical Physics,
Waterloo, ON, Canada
N2L 2W9,\\
Department of Mathematics and Statistics, University of New
Brunswick, Fredericton, NB, Canada E3B 5A3\\} 
\date{\today}
\begin{abstract}

Using a Hamiltonian formulation of the spherically symmetric
gravity-scalar field theory adapted to flat spatial slicing, we
give a construction of the reduced Hamiltonian operator. This
Hamiltonian, together with the null expansion operators presented
in an earlier work, form a framework for studying gravitational
collapse in quantum gravity. We describe a setting for its
numerical implementation, and discuss some conceptual issues
associated with quantum dynamics in a partial gauge fixing.

\end{abstract}

\maketitle

\section{Introduction}

A complete understanding of the many puzzles related to black
holes, such as entropy \cite{entropy}, Hawking radiation
\cite{hawk}, and the final state question \cite{finals}, will
require a quantum theory describing gravitational collapse, where
both matter and geometry are fully dynamical. The ideal would be
the ability to follow the evolution of an initial matter-geometry
quantum state to "black hole formation" and beyond. This goal has
been implicit in many works on black hole physics beginning with
Unruh thirty years ago \cite{unruh}, followed a few years later by
work of Hajicek \cite{hajicek}. More recently it has been
emphasized by Isham as one of the motivations for studying quantum
gravity \cite{isham}. To date, however, no complete quantum
framework is available for studying gravitational collapse of a
scalar field.

This paper continues a recent line of work aimed at developing the
tools necessary to realize this ideal. So far, we have developed
the quantum kinematics for spherically symmetric gravitational
systems, in a setting which leads to a  result concerning the
resolution of the classical singularity at the quantum level
\cite{HWsing}. Furthermore, we have proposed how to define a black
hole at the quantum level, without fixing any classical "horizon
boundaries" \cite{HWdef}. Finally, we formulated the classical
dynamics of black holes in a framework that results in a true
Hamiltonian, rather than a system with a Hamiltonian constraint
\cite{HWclass}.

The current paper is concerned with the development of the quantum
dynamics of the gravity-scalar field system in spherical symmetry.
It consists of two main parts. The first one deals with the
conceptual issues connected with the notion of true dynamics. This
is necessitated by the fact that our treatment of the dynamics is
based neither on Dirac's constraint quantization, nor on a
complete gauge-fixing, but lies in between those two extremes by
employing a partial gauge-fixing only of time. As this type of
approach has - to the best of our knowledge - not been discussed
before in any detail, we include a careful discussion. The second
part is of a more technical nature and describes the construction
of the Hamiltonian operator starting from the reduced Hamiltonian
derived in \cite{HWclass}.

The work is an attempt to complete a proposal for the quantum
theory of the gravity-scalar field system in spherical symmetry.
It provides a calculational framework for studying gravitational
collapse in quantum gravity, which we hope will lead to a better
understanding of Hawking radiation, and of the final state of
gravitational collapse.

The next section discusses some conceptual issues concerning
quantum evolution in a setting where only a partial (time) gauge
fixing is utilized. This is followed in Section III by a review of
the kinematical framework for quantization of the 1+1 dimensional
field theory that describes the gravity-scalar field system.
Section IV contains a construction of the Hamiltonian operator.
Since the classical Hamiltonian contains a square root, we
introduce here a new method of defining the corresponding
operator. In Section V we give the action of the Hamiltonian on
basis states, describe how to implement unitary evolution, and
discuss the issue of dynamical singularity resolution. We conclude
in Section V with a summary and outlook for numerical
implementation.

\section{Conceptual setting}

There are at least two methods to introduce a notion of time into
a theory with time reparameterization invariance. The most obvious
way is by means of gauge-fixing, where  a suitable function on the
classical phase space is taken to be the time function. Another is
the closely related method of using partial and complete
observables \cite{Rov,BD}, where a  degree of freedom is chosen as
the reference clock, and the evolution of the remaining degrees of
freedom is measured with respect to that clock.

We have chosen the former since it is closest to the setting in
which the well-known semiclassical results about black holes have
been derived. In earlier work \cite{HWclass} on the classical
theory, we derived a reduced Hamiltonian for the gravity-scalar
field theory in spherical symmetry by performing only a time gauge
fixing, and leaving the remaining spatial coordinate freedom
untouched. The Hamiltonian constraint was solved as a strong
condition, and a reduced spatial diffeomorphism generator remained
as the only (first class) constraint. Together with the surface
term, this constraint forms the total Hamiltonian. If the surface
term is written as a bulk integral and combined with the other
bulk terms, one can identify the true local Hamiltonian density
and a diffeomorphism generator for the remaining degrees of
freedom.

Fixing only a time gauge raises the question of the extent to
which unambiguous evolution can be achieved in the quantum
dynamics, since evolution still contains a gauge part manifested
in the freedom in the shift function. This may be seen
schematically through the requirement that a gauge condition such
as $f(p,q)=0$ be preserved in time. It leads to the equation
\be 0 = \dot{f} = \{f, \int \left(NH + N^aH_a\right) \}, \ee
which fixes the lapse function $N$ in terms of $N^a$. The latter
remains arbitrary, and must be specified to compute evolution
classically or quantum mechanically. This situation is similar to
Yang-Mills theory where the total Hamiltonian density before gauge
fixing and solving the Gauss law is
\be H_{YM} = \frac{1}{2} (E^2 + B^2) + \Lambda^i G_i, \ee
where $\Lambda^i$ is the Lagrange multiplier and $G^i$ is the
Gauss law expression.

In contrast the Hamiltonian for our problem (in a fixed time
gauge) contains an additional twist (see below for the details).
Schematically, it is of the form
\be H_{grav} = f\left((N^r)';q,\pi\right) + N^r C_r(q,\pi), \ee
where $C_r$ is the radial diffeomorphism generator, $q,\pi$ denote
the collection of canonical phase space variables, and $N^r$ is a
lagrange multiplier (the remaining radial component of the shift
function in spherical symmetry). The similarity between the two
cases (YM and gravity) is the clear separation of the gauge
generators from the "true" Hamiltonians. The difference is that
the gravity case contains Lagrange multiplier dependence also in
the first term, but only through its radial derivative.

This poses the conceptual issue of obtaining unambiguous
evolution, since the gravity reduced Hamiltonian generates a
family of time evolutions for observers specified by $N^r$. This
freedom is limited to the spatial reference system only. Once it
is fixed so is the clock. This situation lies in the middle of the
two common scenarios in dealing with Hamiltonian gravity, where
either no gauges are fixed, or all are fully fixed.

We will address this issue in two steps. Firstly, we obtain an
Hamiltonian operator that depends on $N^r$, determine its action
on basis states, and thereby obtain a prescription for evolution
with any $N^r$ (with the prescribed asymptotic conditions). This
gives a family of evolved states parameterized by $N^r$. A unique
evolution is then obtained by specifying a specific functional
form for this function. This may be viewed as the quantum analog
of evolving with a fixed shift function.

Secondly, we will not impose the diffeomorphism generator as a
constraint on states a'la Dirac. Rather we will give prescriptions
for obtaining information pertaining to the collapse problem in a
manner which is manifestly invariant under this symmetry. This
means looking at the dynamics of certain phase space observables,
such as null expansion, as functions of other phase space
variables. Suitable semiclassical states that are peaked on
classical solutions of the radial diffeormorphism constraint will
be used as the physical states.

Taken together, the implementation of these ideas, provided in the
following  sections, appears to provide a tractable approach to
the gravitational collapse problem in quantum gravity.

\section{Review of the kinematical setting}

Before addressing the issue of quantum dynamics, we briefly review
the quantum kinematical setting, as introduced in \cite{HWsing}.
The starting point is a classical field theory in 1+1 dimensions,
characterized by the canonical pairs $(R,P_R)$, describing the
geometry degrees of freedom, and $(\phi,P_{\phi})$, describing a
scalar field. The basic variables that are turned into quantum
operators are the smeared fields
\be R_f = \int_0^\infty R f\ dr \label{Rf} \ee
and
\be \phi_g = \int_0^\infty \phi g\ dr, \ee
where $f$ and $g$ are suitable test functions. These configuration
variables, together with the translation generators
\be U_\lambda(P_R)(r) \equiv e^{i\lambda P_R(r)} \label{U} \ee
and
\be U_\lambda(P_\phi)(r) \equiv e^{i\lambda P_\phi(r)}, \ee
form a closed Poisson algebra, and are taken as the basic
variables to be converted into operators. All other operators are
be constructed in terms of these.

 The Hilbert space for the
quantum theory is spanned by basis states of the form
\bea &&|e^{i \sum_k a_k P_R(x_k)};e^{i L^2\sum_l b_l
P_\phi(y_l)} \rangle \nn\\
&&\equiv |a_1\ldots a_{N_1};b_1\ldots b_{N_2}\rangle, \eea
where $a_k,b_l$ are real numbers, and $N_1$ and $N_2$ are positive
integers. The factors of $L$ in the exponents reflect the length
dimensions of the respective field variables. The intuitive
picture is that each basis state `tests' the quantum scalar field
at $N$ points in space and the basis of the `excitation space' at
each point consists of (generalized) plane waves. The inner
product is
\bea \langle a_1\ldots a_{N_1};b_1 \ldots b_{N_2} |a_1'\ldots
a_{N_1}';b_1'\ldots b_{N_2}'\rangle \nn\\
&& = \delta_{a_1,a_1'} \ldots \delta_{b_{N_2},b'_{N_2}}, \eea
if the states contain the same number of sampled points, and is
zero otherwise. The action of the basic operators is
\bea &&\hat{R}_f\
|a_1\ldots a_{N_1};b_1\ldots b_{N_2}\rangle = \nn\\
&&L^2 \sum_k a_k f(x_k)|a_1\ldots a_{N_1};b_1\ldots b_{N_2}\rangle
\nn\\
&&\hat{\phi}_g\
|a_1\ldots a_{N_1};b_1\ldots b_{N_2}\rangle = \nn\\
&&L^2 \sum_l b_l g(y_l)|a_1 \ldots a_{N_1};b_1 \ldots
b_{N_2}\rangle \eea
and
\bea && \widehat{e^{i \lambda_j P_R(x_j)}}|a_1 \ldots
a_{N_1};b_1\ldots b_{N_2}\rangle
 = \nn\\
&&|a_1\ldots, a_j-\lambda_j,\ldots
a_{N_1};b_1\ldots b_{N_2}\rangle \nn\\
&& \widehat{e^{i \mu_k P_{\phi}(y_k)}}|a_1\ldots a_{N_1};b_1\ldots
b_{N_2}\rangle
 = \nn\\
&&|a_1\ldots a_{N_1};b_1\ldots,b_k-\mu_k,\ldots b_{N_2}\rangle
\eea
where $a_j$ (resp. $b_k$) is $0$ if the point $x_j$ (resp. $y_k$)
is not part of the original basis state. In this case the action
creates a new `excitation' at the point $x_j$ (resp. $y_k$) with
`mode' $-\lambda_j$ (resp. $-\mu_k$). These definitions give the
commutator
\be \left[ \hat{R}_f, \widehat{e^{i\lambda P_R(r)}} \right] =
-\lambda f(x) L^2 \widehat{e^{i\lambda P_R(r)}}.
\ee
Comparing with the classical Poisson bracket relation, and using
the Poisson bracket commutator correspondence, it turns out that
$L=\sqrt{2} l_P$, where $l_P$ is the Planck length. A similar
commutator relation holds for the matter scalar field.

In this formalism we can also define other operators of interest.
For example for the Hamiltonian, we need operator analogs of the
radial derivatives $R'$ and $\phi'$, and the square of the
momentum $P_\phi^2$.

 Definitions for the operators corresponding to $R'$ and other
derivatives of fields are obtained by implementing the idea of
finite differencing using the operator $\hat{R}_f$ (\ref{Rf}). We
use narrow Gaussians with variance proportional to the Planck
scale, peaked at coordinate points $r_k+\epsilon l_P$, where
$0<\epsilon \ll 1$ is a parameter designed to sample neighbouring
points\footnote{A possible choice is to use characteristic
functions of bounded subsets $A$ of the real line. This is the
easiest way to localize the configuration fields around points of
interest,  and to construct operators corresponding to the
derivatives of field operators. It offers an alternative to the
construction in \cite{HWdef}.}:
\be
f_{\epsilon}(r,r_k)= \frac{1}{\sqrt{2\pi}}\ {\rm exp} \left[-\
\frac{(r-r_k-\epsilon l_P)^2} {2 l_P^2}\right]
\ee
Denoting $R_{f_\epsilon}$ by $R_{\epsilon}$ for this class of test
functions we define
\be
  \hat{R'}(r_k):= \frac{1}{l_P \epsilon}\ \left( \hat{R}_{\epsilon} - \hat{R}_{0}
\right).
\label{R'}
\ee
A further motivation of this form is that in the gauge $R=r$ the
corresponding classical expression gives unity in the limit
$\epsilon \rightarrow 0$. This definition captures the simplest
finite difference approximation to the derivative at the operator
level. Second derivatives may be similarly defined by converting a
finite differencing scheme into the corresponding operator.

The quantization defined above does not give definitions for
operators corresponding to momenta, since only translation
generators are directly realized as operators at the first step.
This of course is similar to any quantum theory on a lattice.
Operators for momenta can however be realized using the
translation generators, for example by expressions such as
\be
  \hat{P}_\phi^\lambda := \frac{l_P}{2i\lambda}\ \left( \hat{U}_\lambda -
\hat{U}_\lambda^\dagger\right)
\ee
and
\be
\hat{P}_\phi^2 :=  \frac{l_P^2}{\lambda^2}\ \left(
2-\hat{U}_\lambda - \hat{U}_\lambda^\dagger \right)
\ee
These $\lambda$ dependent expressions will be utilized below for
the Hamiltonian operator.

 Finally, let us note that the background independent (ie.
metric free) quantization outlined here is not the same as the one
for the scalar field reported in \cite{als,l}, where the basic
variables used are the integral of the scalar field momentum over
space $\int_\Sigma P_\phi$, and the exponential of the scalar
configuration $e^{i\lambda \phi}$. The present approach has the
advantage that it is fairly straightforward to write a local
Hamiltonian density operator using the translation and $\phi'$
operators. A further technical point of difference, apart from the
choice of basic variables, arises from the fact that $R$ and
$\phi$ are scalars, and $P_R$ and $P_\phi$ are scalar densities of
weight 1. This means that the functions $f$ in (\ref{Rf}) and the
$\lambda$ in (\ref{U}) are scalar densities of weights $1$ and
$-1$ respectively. These combine to give eigenvalues of
$\hat{R}_f$ and $\hat{\phi}_f$ that are scalars. In contrast, for
the quantization studied in \cite{als,l}, it is the
space-integrated momentum density that is diagonal; this is the
direct analog of the surface integrated densitized dreibein in
loop quantum gravity.

 This ends our review of the kinematical
setup. Our goal in the next section is to define the Hamiltonian
operator for the collapse problem on this Hilbert space.

\section{Hamiltonian}

The classical Hamiltonian in a time gauge fixing corresponding to
flat spatial slicing given was derived in \cite{HWclass}. The
classical phase space before gauge fixing has an extra pair of
canonical variables $(\Lambda, P_\Lambda)$, with the spatial
metric
\be ds^2 = \Lambda^2 dr^2 + R^2 d\Omega^2. \ee
The condition $\Lambda = 1$ is second class with the Hamiltonian
constraint, which is solved classically for the conjugate momentum
$P_\Lambda$. This leads to the reduced Hamiltonian
\bea
H_{R}^G &=& \int_0^\infty \left[(N^r)'P_\Lambda + N^r(P_RR' +
P_\phi \phi')\right]dr
\nn\\
&=&\int_0^\infty (N^r)'\left( R P_R + \sqrt{(R P_R)^2 - X}\right) dr\nn\\
  && + \int_0^\infty N^r(P_RR' + P_\phi \phi')\ dr,
\label{Hred}
\eea
where $N^r$ is the remaining (and still arbitrary) non-zero
component of the shift function $N^a$ after imposing spherical
symmetry,
\be
X =  16R^2 (2RR'' - 1 + R'^2) + 16R^2 H_\phi, \label{X}
\ee
and
\be
 H_\phi = \frac{P_\phi^2}{2R^2} + \frac{R^2}{2}\ \phi'^2.
\ee
The Hamiltonian (\ref{Hred}) is obtained by writing the surface
term in the reduced action as a bulk integral and combining terms.
It gives the time-gauge fixed evolution equations for the fields
$R(r,t)$ and $\phi(r,t)$, and their canonical conjugates. Our main
goal in this paper is to construct the corresponding operator.

Let us write the first part of the Hamiltonian density (excluding
the radial diffeomorphism term) as
\be H_R^G = h_1 + \sqrt{h_1^2 -h_2^2 -h_3^2 + {\cal R}} \equiv h_1
+ A \ee
where we have defined
\be h_1 = P_R R, \ \ \ \ \ \, {\cal R} = -16R^2 (2RR'' - 1 + R'^2)
\label{h1R} \ee
\be h_2^2 = 8P_\phi^2, \ \ \ \ \ \ h_3^2 = 8R^4\phi'^2.
\label{h2h3} \ee

There are two ways to construct an operator corresponding to a
square root. One approach is to find the eigenvalues of its
argument and work with the corresponding basis of eigenvectors.
The operator can then be defined as the square root of the
eigenvalues if the spectrum is positive semi-definite.  A
potential alternative is to see if the square root operator can be
defined using Dirac's idea.

In this procedure an operator for the classical function
$H=\sqrt{p_i^2 + m^2}$ $(i=1\cdots 3)$ is constructed by writing
$\hat{H}= \alpha_i\hat{p}_i + \beta m$ using anticommuting
matrices $\alpha_i,\beta$, such that
$\hat{H}^2=(\hat{p}_i^2+m^2)I$, where $H^2$ defined by matrix
multiplication. This works  because the momentum components
$\hat{p}_i$ commute. This is not true for the elements $h_i$ in
the argument of our Hamiltonian, so Dirac's trick cannot be used,
at least in its basic form.

There is the additional problem that the term ${\cal R}$, which is
the Ricci scalar, is not a squared quantity, and so can be
negative. (In the gauge $R=r$ this term vanishes since we are
using the flat slice asymptotic conditions \cite{HWclass}.)
$\hat{\cal R}$ is however diagonal in the basis since the operator
analogs of the fields $R$, $R'$ and $R''$ are all diagonal
\cite{HWdef}. Therefore at least for the subset of states where
its eigenvalues $\lambda_{\cal R}$ are positive we can define an
operator $\hat{h}_4$ whose eigenvalues are  $\sqrt{\lambda_{\cal
R}}$. We will return to this issue below after describing an
approach to address the square root problem. For now, and the
following discussion, we define
\be h_4^2:={\cal R}. \ee

The idea is to obtain a definition of the square root part
$\hat{A}$ of the Hamiltonian operator by working in a larger
Hilbert space
\be {\cal H} = {\cal H}_{Kin}\otimes V, \ee
for some $V$ to be specified, and writing
\be \hat{A} = \hat{h}_k \bar{e}^k, \label{A1} \ee
where the $\hat{h}_k$ act in ${\cal H}_{Kin}$, and the $\bar{e}^k$
act in $V$. The key requirement is that the action of $\hat{A}$ in
${\cal H}$  must be such that
\be \hat{A}^2  = \hat{h}_k \hat{h}_l\ \eta^{kl} I, \label{cond}
\ee
where $\eta^{kl}= {\rm diag}(+--+)$ and $I$ is the identity
operator in the space $V$. All other operators $\hat{O}$ acting in
${\cal H}_{Kin}$ are extended to ${\cal H}$ by the identity action
in $V$. Since the $h_k$ do not commute, this means that the
operators $\bar{e}^k$ must satisfy
\be \bar{e}^k\bar{e}^l =\eta^{kl}I. \label{cond2} \ee
That no such operators exist for any space $V$ may be seen by the
following argument. Let $|\psi_k\rangle$ be a complete set of
normalisable states, and let us assume that operators satisfying
(\ref{cond2}) exist. Then we are led to contradictions such as
\bea 1 &=& \langle \psi| \psi\rangle = \langle \psi|
\left(\bar{e}^1\right)^2 |\psi\rangle \nn\\
&=& \langle\psi|\bar{e}^1 (\bar{e}^4)^2 \bar{e}^1 \psi\rangle
 = \langle \psi|\bar{e}^1 \bar{e}^4\sum_k |\psi_k\rangle\langle\psi_k|
\bar{e}^4\bar{e}^1 |\psi\rangle =0.
 \eea

There is a way out of this situation if we demand the weaker
condition that
\be \langle \Psi|\hat{A}^2|\Psi \rangle = \langle \Psi |\hat{h}_k
\hat{h}_l\ \eta^{kl}| \Psi\rangle, \label{expec} \ee
for every $|\Psi\rangle \in {\cal H}$ that is of a form specified
below. The basic idea is to introduce a four-dimensional Euclidean
vector space $V$ and consider its decomposition into orthogonal
subspaces
\be V=\oplus_k V_k \ee
$k=1\cdots 4$. Let us denote by $P_k$ the projection operators
onto these subspaces. By definition the $P_k$ satisfy $P_kP_l =
P_k \delta_{kl}$. Now define operators $\bar{e}_k$ by
\be \bar{e}_k = P_k \ee
for $k=1,4$, and
\be \bar{e}_k = iP_k \ee
for $k=2,3$. These give
\be \hat{A}^2 = \hat{h}_k\hat{h}_l \bar{e}_k \bar{e}_l =
\hat{h}_k\hat{h}_l P_k \eta_{kl} \ee
Consider now the action of $\hat{A}^2$ on states $|\Psi\rangle$ of
the form
\be |\Psi\rangle  \equiv |\psi\rangle |\rho\rangle \ee
where $|\psi\rangle \in {\cal H}_{Kin}$,
\be |\rho\rangle =\sum_k|\rho_k\rangle, \ee
and the $|\rho_k\rangle$ denote the basis of V corresponding to
the decomposition into the $V_k$. The result is
\bea \hat{A}^2|\Psi\rangle
&=&\sum_{k,l}\left(\hat{h}_l\hat{h}_k|\psi\rangle\right)\ P_l
|\rho_k\rangle \nn \\
&=&
\sum_{k}\left(\hat{h}_k\hat{h}_k|\psi\rangle\right)|\rho_k\rangle.
\label{H2psi} \eea
From this it is evident that (\ref{expec}) holds. Finally, we
still need a  refinement to get a self-adjoint $\hat{A}$. To do
this we define the operators $\hat{h}_2$ and $\hat{h}_3$ such that
they have the property
\be \hat{h}_2^\dagger = \hat{h}_2\ \ \ \ \ \hat{h}_3^\dagger =
\hat{h}_3. \ee
These ensure that the square root operator $\hat{A}$ in the
extended Hilbert space is self-adjoint.

The last issue to address is the definition of the operator
corresponding to ${\cal R}$ (the Ricci scalar) in Eqn.
(\ref{h1R}). This contains derivatives of $R$ which are defined by
the finite difference operators introduced in \cite{HWdef}. Since
these are diagonal on basis states, so is $\hat{\cal R}$ . Now, in
the gauge $R=r$, which fixes the shift vector $N^r$ to be
proportional to $1/\sqrt{r}$ everywhere, the Ricci scalar
vanishes. Our Hamiltonian is not in this coordinate gauge (which
is still free), but as alluded to in Sec. II, the idea is to
evolve quantum states with this choice of lapse function, with
initial states for which the values $a_k$ of the radial field $R$
are distributed linearly with the graph points $r_k$, ie. $a_k
\sim r_k$. This guarantees that the eigenvalue of $\hat{\cal R}$
vanishes, at least initially. However it will not remain so
because the Hamiltonian contains terms with $P_R$, which is
represented by the operator
\be \hat{P}_R(r_k) :=  \frac{1}{2i\lambda}\ \left( e^{i\lambda
P_R(r_k)} - e^{-i\lambda P_R(r_k)}\right). \label{PR} \ee
The translation operators on the r.h.s. act to shift excitations
at the point $r_k$ which result in the eigenvalue $\lambda_{\cal
R}$ of $\hat{\cal R}$ being moved from zero at the selected point.
However this move is very small since $0<\lambda <<1$. Because of
this we define the action of $\hat{h}_4$ by
\be \hat{h}_4 |\psi\rangle = \sqrt{|\lambda_{\cal
R}|}|\psi\rangle, \ee
where  $|\psi\rangle$ is a basis state. This definition makes the
assumption that, starting from a basis state with $\lambda_{\cal
R} =0$, the deviation from zero after the action of $\hat{P}_R$ is
positive.

This completes our prescription for the Hamiltonian operator for
the scalar field collapse problem. With each of its constituents
well defined, it is straightforward to compute its action on a
basis state. The fact that this can be done in contrast to the
case for full gauge fixing \cite{finals} represents some progress
which we anticipate will lead to concrete calculations for quantum
collapse. A strategy to do this is presented in the next section.

We close with a discussion of the quantization choices, or
ambiguities, inherent in the steps outlined here. Firstly, let us
note that the general procedure used in obtaining the square root
operator is new. The only other possible approach is to seek the
spectrum of the argument of the square root, which given its form
appears a formidable task. If this could be done, it would
represent another, possibly physically distinct, choice of
Hamiltonian for this problem. This circumstance would be analogous
to the Dirac and Klein-Gordon Hamiltonians. Secondly, $h_1$
contains products of non-commuting operators so a choice must be
made for its operator version. The natural one is to take the
symmetric product $(RP_R + P_RR)/2$. Finally, the only other
ambiguity in the definition of the Hamiltonian is in the choice of
"lattice" parameter $\lambda$. It is natural to also use the same
parameter in the Gaussian smearing functions used in operators
such as $\hat{R}_f$. In implementing evolution numerically, the
goal of course is to ensure that physical results do not depend on
this parameter. Like any numerical computation, one would like to
see that the evolution of a fixed initial state leads to a
convergent answer for the final state after a fixed number of time
steps. This means that when the computation is repeated for
successively smaller values of $\lambda$, the answers for evolved
physical variables have asymptotic "continuum" values.

\section{Quantum evolution}
In classical numerical simulations of scalar field collapse the
general approach is to start with initial data representing a
shell of scalar field, and to evolve it using some choice of lapse
and shift functions \cite{sccoll}. At each step of the simulation
an "apparent horizon" check is made as a criterion for black hole
formation. This is a null geodesic trapping condition at each step
or leaf of the evolution. In spherical symmetry the goal is to
find the outermost radial location on each leaf where the
condition is satisfied. The evolution of this location is taken to
represent the dynamical boundary or horizon of the black hole.

There are at least two concrete versions of what is a dynamical
horizon. The first was formulated by Hayward \cite{hay}. In
addition to the usual criteria for null geodesic expansions, this
work consists of additional conditions designed to distinguish
future, past, inner and outer local horizons. The second
\cite{ash} lifts some of these conditions, and points out that the
resulting definition of dynamical horizon allows a nice
formulation of local flux laws. The common and minimal feature of
both definitions are the conditions
\be \theta_+ =0, \ \ \ \ \ \  \theta_-<0 \ee
where $\theta_\pm $ are the in(out)going null geodesic expansions.
Equivalent information is captured in the observable
$\theta_+\theta_-$, which goes from negative to positive as a
dynamical black hole boundary is crossed.

With the Hamiltonian operator defined in the last section, and the
operator analogs of the null expansion operators given in
\cite{HWdef}, we are in a position to give a procedure for a
quantum collapse calculation. This involves specifying (i) initial
states, (ii) an evolution procedure, and (iii) a quantum test for
black hole formation.

The first question is what are suitable initial states. The basis
states represent values of the scalar field $\phi(r,t)$ and the
radial field $R(r,t)$ at a set of discrete coordinate points
$r_1\cdots r_k$, which is a sample of the half line. As such these
states may be compared with the discrete data for a classical
numerical simulation. This suggests the use of "profile" states
where we take for example the scalar field excitations to have a
gaussian profile, and the radial excitations to be linearly
distributed. Another possibility is the use of suitable coherent
\cite{tw} or other form of semi-classical states that are peaked
at classical configurations. These would be infinite linear
combinations of basis states, and so computation with them would
be more involved, especially in the present field theory setting.

The second step is the implementation of evolution. Rather than
constructing a finite evolution operator by exponentiation of the
Hamiltonian, which is very cumbersome, it is more suitable to
implement repeated infinitesimal evolution using a suitable
scheme. The simplest possibility
\be |\psi\rangle_{t+\Delta t} = (I - i\Delta t \hat{H})|\psi
\rangle_t \ee
is not unitary. However there are unitary schemes available for
this purpose. One example is based on the well known
Crank-Nicholson method, where the Schrodinger equation in discrete
time (labelled by $n$) is written as
\be \frac{i}{\Delta t}\ \left(|\psi \rangle_{n+1} -
|\psi\rangle_n\right) = \frac{1}{2}\ \left(
\hat{H}|\psi\rangle_{n+1} + \hat{H}|\psi\rangle_n \right).
\ee
This leads to the manifestly unitary (but implicit) evolution
scheme given by
\be
\left(1+\frac{i}{2}\ \Delta t\ \hat{H} \right)
|\psi\rangle_{n+1} = \left(1-\frac{i}{2}\ \Delta t\ \hat{H}
\right) |\psi\rangle_n
\ee
  Since $\hat{H}$ depends also on the free (classical)
function $N^r$, so does the evolved state $|\psi\rangle_{n+1}$.
This function must be fixed (with the fall off condition $N^r\sim
1/\sqrt{r}$) to get a unique evolution \cite{HWclass}. The
simplest choice is to take this form for all points $r_k$ in the
chosen initial state.

The third step is to implement a test for black hole formation at
each time step of the evolution. As mentioned above, the minimum
requirement for this is that we must have operators corresponding
to the null expansions. A prescription for constructing these were
given in \cite{HWdef}. In addition we require this test to be
invariant under the remaining radial diffeomorphism constraint.
This is achieved, for example, by looking at the quantities
$\langle\theta_+(r_k,t)\theta_-(r_k,t)\rangle$ as functions of
$<R(r_k,t)>$, where the expectation values are in the state
arrived at by stepwise evolution from some initial state. The
resulting curve is radial diffeomorphism invariant, and its
intersection with the $\langle\hat{R}\rangle$ axis gives the
location and size of the evolving horizon. It is the dynamics of
this curve which is of interest for black hole formation and
subsequent evolution at the quantum level.

There are many other useful diffeomorphism invariant quantities of
interest that can be computed at each time step. Two examples are
the scalar field configuration $\langle\phi(r_k,t)\rangle$, and a
curvature measure such as $\langle \hat{\pi}(r_k,t)\rangle$ (the
trace of the ADM momentum)\cite{HWsing}, both viewed as functions
of $\langle \hat{R}(r_k,t)\rangle$.

The second and third steps are to be repeated for multiple time
steps to extract time dependent profiles of the functions of
interest, such as the ones just mentioned. It is in this manner
that physical information about collapse may be obtained in the
kinematical Hilbert space after fixing the time gauge classically.
Although evolution is with respect to a fixed $N^r$ so that the
evolved states depend on it, the physical information contained in
the suggested functions is independent of the coordinate points
$r_k$: radial diffeormorphisms act to shift the points $r_k$ to
$r_k'$ without changing the values of the field variables.

\section{Dynamical avoidance of the singularity}

In earlier work we gave a construction of an operator
corresponding to the classical variable $1/R$ that is bounded on
the Hilbert space we are using for the quantum theory
\cite{HWsing}. This result has the direct consequence that
curvature singularities in spherical symmetry are avoided. This is
because phase space variables that classically diverge at the
singularity do so as positive powers of $1/R$. The corresponding
quantum operators are constructed as products of the $1/R$
operators, and so are also bounded. This does not involve the
Hamiltonian in any way so the result may be viewed as kinematical
singularity avoidance. In this section we show, using the
construction of the Hamiltonian operator given above, that
inclusion of dynamics does not alter this result.

The fundamental question here is whether quantum evolution remains
well-defined through the region of highest curvature, or whether
it stalls or breaks down there. Classically dynamics is encoded
either in (i) the Hamiltonian constraint, or (ii) a Hamiltonian
derived from a partial (time) gauge-fixing, or (iii) a Hamiltonian
derived from a complete gauge-fixing as in \cite{unruh}. Although
our work is concerned with the second case, we will look at the
issue of singularity avoidance from all three points of view.
Let us consider first the Hamiltonian constraint. Its classical
expression is
\bea H &=& \frac{1}{R^2\Lambda}\left[\frac{1}{8}\ (P_\Lambda
\Lambda)^2 - \frac{1}{4}(P_\Lambda \Lambda)(P_R R)\right]
\nn\\
&&+ \frac{2}{\Lambda^2}\left[ 2RR''\Lambda -2RR'\Lambda' -
\Lambda^3
+\Lambda R'^2 \right]\nn\\
&& + \left[\frac{P_\phi^2}{2\Lambda R^2} + \frac{R^2}{2\Lambda}\
\phi'^2 \right]. \eea
Due to the $1/R$ factors, it is divergent at the points $r$ where
$R=0$. Upon quantization these factors turn into bounded operators
as shown in \cite{HWsing}. From this result, and the form of the
basic quantum operators it is clear that the quantum Hamiltonian
constraint has finite action on all states, including those for
which the eigenvalue of $\hat{R}_f$ is zero. The action of the
constraint operator on a state with maximum eigenvalue of the
$1/R$ operator gives a linear combination of basis states that
contain shifts in excitation values of the fields generated by the
action of the corresponding momentum operators (\ref{PR}). It is
not difficult to see that due to this, each of the states in the
linear combination correspond to a lower eigenvalue of the $1/R$
operator. Thus in this sense a time step "evolution" by the action
of the Hamiltonian constraint  on a state of maximum value of
curvature gives a new state in which its expectation value is
lower than the maximum.

Consider next the fully reduced Hamiltonian that was found in
\cite{unruh}
\be
 H_S = \int dr \left[
\frac{P_{\phi}^2}{4r^2} + r^2 \phi^{'2} \right] exp \left(
\int_{\infty}^r S_{\phi} (r') dr' \right)\, \ee
where
\be S_{\phi} (r) = \frac{P_{\phi}^2}{8r^3} + \frac{r\phi^{'2}}{2}.
\ee
This is clearly divergent at $r=0$. As $r$ is a parameter rather
than a configuration field variable, quantization of the totally
reduced theory cannot resolve the singularity at the quantum
level.

Finally, let us consider the Hamiltonian whose quantization is the
subject of this paper. Firstly, an inspection of the classical
reduced Hamiltonian (\ref{Hred}) shows that, unlike the
Hamiltonian constraint, it has no manifestly divergent $1/R$
factors. This surprising feature is just a consequence of the time
gauge fixing, and concomitant strong solution of the Hamiltonian
constraint. (If one continues the reduction process further by the
gauge choice $R=r$, the divergence reappears via the shift
function, which is proportional to  $1/\sqrt{r}$ \cite{HWclass}.
This becomes the source of the divergence at $r=0$ in the fully
reduced Hamiltonian.) Secondly, the Hamiltonian operator  has well
defined action on basis states, and evolution does not stall on
states with maximum eigenvalue of the $1/R$ operator. Rather, it
gives a linear combination of states each of which has a lower
eigenvalue of this operator (for the same reason as for the action
of the Hamiltonian constraint discussed above).
In closing this section, we compare the above scenario for
dynamical singularity resolution with the cosmological case
studied in \cite{abl,HWcos}, and the Schwarzschild black hole
studied in \cite{ab}. In both these cases the systems are finite
dimensional, unlike the model in this paper. In the first case,
the action of the Hamiltonian constraint on a basis state gives a
finite difference equation with coefficients such that action on
the state of zero volume gives a bounce. In the second case the
interior (Kantowksi-Sachs) and exterior of the Schwarzschild
spacetimes are quantized with appropriate matching at the event
horizon, which is taken as the fundamental classical dividing
line. Our perspective is that these cases are quite different from
the matter coupled case we treat here. In particular, with
non-vanishing matter fields, the extended Schwarzschild spacetime
does not arise, so quantizing it to discuss singularity resolution
is not relevant for the quantum treatment of the  collapse
problem.

\section{Discussion}

We have constructed the quantum Hamiltonian for the spherically
symmetric system of gravity coupled to a scalar field in a fixed
time gauge. Together with earlier papers
\cite{HWsing,HWdef,HWclass}, this work completes the construction
of a quantum theory for studying the gravitational collapse of a
scalar field in spherical symmetry. We are now in a position to
address the questions surrounding black holes that have been
generally acknowledged to find a resolution only within a full
quantum treatment.

Any application of our formalism to a given physical situation
will require a choice of initial state. As all the black holes
that have been detected so far have macroscopic size, the first
task is to determine how to represent what we know as a classical
black hole in the quantum theory. This obviously calls for a
construction of the semiclassical sector, including the search for
suitable semiclassical states. This will be the subject of a
future publication \cite{HWsemicl}. Once the issue of what initial
state to take has been settled, one can then investigate the
quantum time evolution of the system.

Among the problems of physical interest is the evolution of a
quantum state that satisfies the quantum horizon conditions. One
would like to know how this evolution depends on the matter part
of the state, and how the horizon grows or shrinks. An
indispensable tool for such questions will be the null expansion
operators which were constructed in \cite{HWdef}, which serve as
horizon finders. It should be pointed out that, even if one does
not subscribe to the exact definition of the horizon of a black
hole proposed in \cite{HWdef}, these operators would still play an
important role in any other approach to a quantum description of
black hole horizons.

Perhaps the most interesting questions to address  in this
framework are whether and how Hawking radiation arises, and what
is the end point of collapse. A preliminary investigation
\cite{HWessay}, indicates that the end result of gravitational
collapse is a Planck size remnant. This appears to be intimately
connected with the existence of an upper bound on curvature, which
in turn has an appealing intuitive analogy with Fermi pressure;
whereas the latter can be ultimately overcome by gravitational
forces, the former cannot since it is a fundamental quantum
gravity effect. It represents the ultimate limit to which matter
can be compactified. Finally, as time evolution in our setting is
unitary, one can surmise already at this point that the solution
to the so-called information loss paradox is that there has never
been a paradox in the first place.

Apart from applications, another potential line of investigation
is to improve on the framework developed here. One of the
questions here concerns time gauge fixing. There are of course
many other choices, so it would be useful to see if there are
others that might lead to simpler reduced Hamiltonians. One
possibility is to look at only the time gauge fixing used in
\cite{unruh}, without fixing the radial gauge $R=r$. Another
approach to the same issues using the connection-triad variables
is being developed by Bojowald and collaborators \cite{martin}.
This work offers a parallel approach in the loop quantum gravity
programme.

\acknowledgments{This work was supported in part by the Natural
Science and Engineering Research Council of Canada. V.H. thanks
the National Center for Physics, Islamabad, for kind hospitality
where this work was completed.}

\end{document}